# Redwood: A Frontier AI Accelerator Designed, Verified, and Deployed from Scratch in 2 Weeks by AI

Architect Labs Team*
Palo Alto, California

*Abstract*— Modern AI workloads and the hardware that runs them evolve on different timescales: architectural definition precedes volume silicon by years, while target workloads shift in months. Design decisions are therefore committed under deep uncertainty and paid for twice, once in the generality added as a hedge, and again when new workloads map poorly onto frozen silicon. As Moore's Law stagnates, specialization is the main remaining source of performance-per-watt and demands a design cycle that runs at the cadence of the workloads. We present an end-to-end AI system that collapses the software-to-silicon stack into a single optimization loop, where hardware and software are co-designed and verified under one objective. Its first demonstration is Redwood, a frontier AI accelerator built for single-batch, low-power, ultra-low-latency inference for physical AI. From a high-level specification by two human architects, the system autonomously generated the performance model, RTL design, UVM environments, formal proofs, firmware, and kernels in under two weeks with no human intervention below the specification. Every block reached 95% coverage via commercial EDA tools, our proprietary formal engine, and hardware-in-the-loop validation. Specification changes were reverified and redeployed to hardware in under 48 hours. Redwood Nano, its ultra-low-power FPGA variant, runs multi-billion-parameter models like Llama and Qwen. Projected onto Samsung 8 nm, the Jetson Orin Nano's process class, Redwood delivers 1.75x the throughput at 1.9x lower power, a 3.4x performance-per-watt gain against a measured Jetson baseline on the same models. Qwen running on Redwood also helped design next-generation Redwood, an early step toward recursive self-improvement. To our knowledge, this is the first production-worthy AI accelerator designed end-to-end by an AI system and running a modern AI model.

## I. Introduction

The EDA industry now reports order-of-magnitude gains from AI across RTL generation, verification, debugging, and exploration, including claims of reducing multi-month efforts to days and achieving up to 10 × productivity across design and verification workflows [1], [2], [3]. However, only 14% of IC/ASIC projects achieve first-silicon success, the lowest rate in two decades, while 75% are behind schedule as chip programs contend with escalating design complexity, driven by heterogeneous integration, advanced-node physical effects, and increasingly stringent power, performance, and area (PPA) constraints [4], [5]. This divergence highlights a deep misalignment between task-level productivity claims and complete program outcomes: AI has accelerated individual activities, but it has not demonstrated clear end-to-end program improvement on increasingly complex SoCs.

At the same time, publicly demonstrated end-to-end AI-generated designs have remained limited to simple examples, such as toy RISC-V cores or hardened numeric datapaths. Almost none have been validated on physical hardware, which is ultimately the final constraint of hardware design. We argue that the opportunity for AI in hardware design is not task acceleration within the existing flows but the reimagining of the entire flow itself. When architecture, RTL, verification, firmware, and kernels are generated from a single specification and optimized under one objective, the sequential handoffs that dominate program latency disappear, and hardware-software co-design becomes a property of the system rather than a coordination process between teams.

To address this, we introduce Redwood, a frontier AI accelerator designed, verified, programmed, and deployed end-to-end by such a system. Two human architects captured the workload and architectural constraints in a high-level specification. From that specification, the system autonomously generated the performance model, RTL, UVM environments, formal proofs, firmware, drivers, and custom compute kernels. In under two weeks, it produced the complete design from scratch, closed every block at 95% code and functional coverage, and deployed the Redwood Nano configuration on an AMD Versal FPGA. A third week brought Qwen3-0.6B inference online [6]. Each architectural change during this period was regenerated, reverified, and redeployed to hardware in under 48 hours. Evaluated on a Samsung 8 nm-class process comparable to that used by NVIDIA Jetson Orin Nano, Redwood Nano is projected to deliver 1.75x the decode throughput at 1.9x lower power, a 3.4x improvement in performance per watt, against the measured Jetson baseline running the same model.

The remainder of the paper is organized as follows. Section II describes the Redwood architecture. Section III presents the programming model. Section IV evaluates Redwood Nano on FPGA and projects its performance, area, and power. Section V details the Architect Labs AI system that produced the design, including automated verification, microarchitectural exploration, and firmware and kernel generation, and reports an early demonstration of recursive self-improvement of AI model and the hardware that fuels it.

## II. Architecture

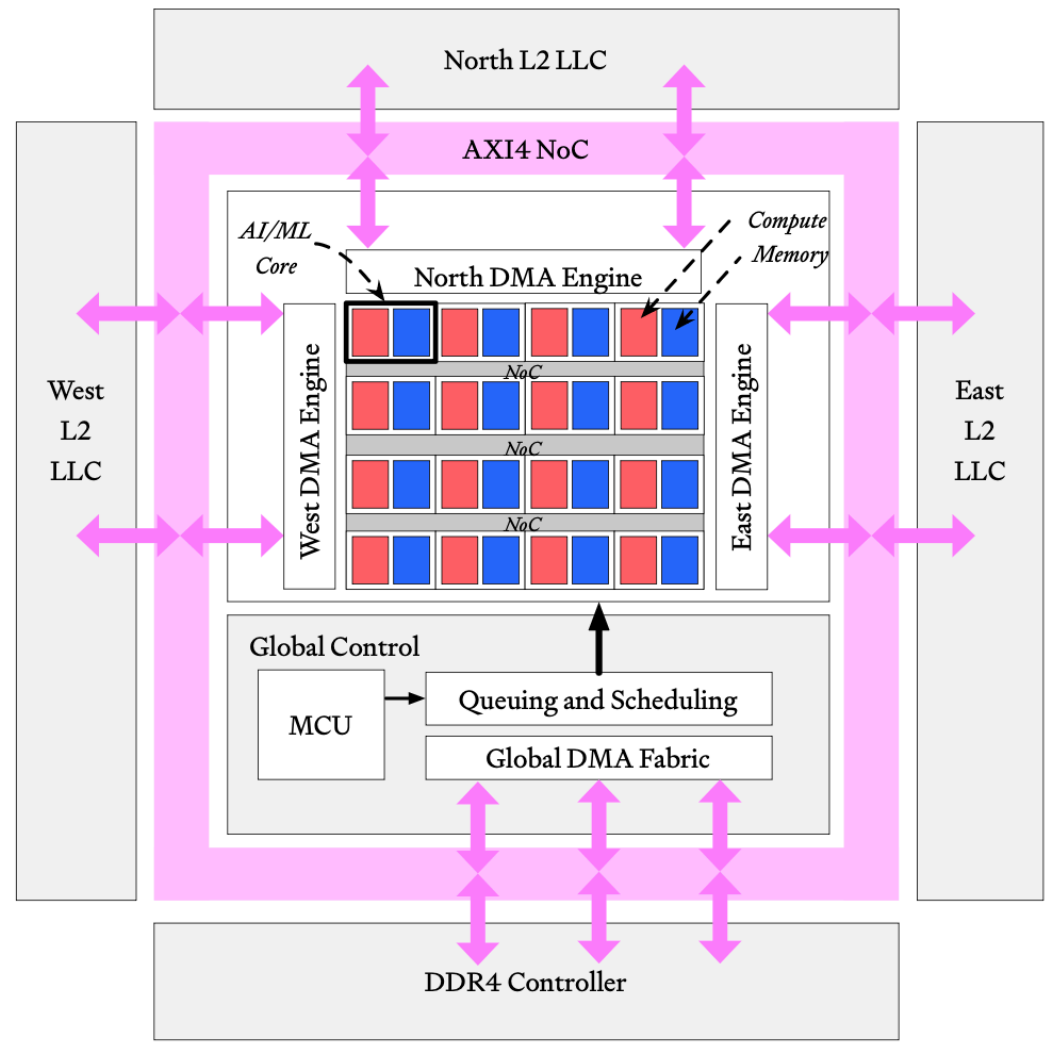

Fig. 1: Redwood SoC architecture.[1]

Redwood is a tile-based, spatial-dataflow accelerator that uses standard AXI4 memory-mapped interfaces: AXI-Lite for control

*The full list of contributors is in the appendix.

[1]The AXI4 NoC (pink) shape here is left abstract (in general not a ring), and it is assumed LLC is much closer and has lower-latency access than the external memory interface (DDR, HBM).

and configuration, and wide AXI4 with full burst support for bulk data (Fig. 1). Dedicated DMA engines move all traffic to and from external DRAM. A global DMA (GDMA) fabric performs bulk memory-to-memory transfers between external memory and the on-chip West, North, and East last-level SRAM banks (LLCs), while edge DMA engines stage data on and off the compute fabric. A global control region sequences the accelerator and contains the global control core (MCU), a global task manager, and a 48-bit global timer (HAC) broadcast to every tile for time-fenced scheduling. This region launches, orchestrates, and cleans up kernels such as FlashAttention [7] and GEMM/GEMV. Because the memory interface is confined to modular DMA engines, Redwood can be integrated within a larger SoC or packaged as a standalone chiplet. The DMA back ends can also be retargeted from AXI4 to protocols such as ACE and CHI without disturbing the compute fabric.

The compute fabric is an $N \times M$ mesh of identical tiles surrounded by edge DMA engines. Each tile pairs a RISC-V-based tile control core (CRV) with compute engines co-designed for transformer inference. The matrix engine (CMXM) provides systolic GEMM and matrix-vector (GEMV) datapaths and streams directly into the vector engine (CVXM), which provides SIMD, transpose, and floating-point activation units. Wide, banked scratchpad memories minimize data movement within Redwood. The compute engines are co-designed with the kernel software, so they map directly onto dominant transformer operators—attention, GEMM, normalization, and activations—and execute kernels such as FlashAttention and GEMM as hardware-scheduled tasks rather than generic instruction streams. A high-bandwidth, internally designed, credit-based NoC carries tile-to-tile, DMA-to-tile, and tile-to-DMA traffic. It provides low-overhead broadcast and multicast, table-based stream redirection, and per-link flow control.

### A. Tile Architecture

Each tile within the Redwood fabric is divided into a front end (FE) and a back end (BE), as shown in Fig. 2. The FE handles control and programming, while the BE handles data movement and computation. Separating sparse control from high-bandwidth data processing allows the FE to operate in a slower clock domain and, in some cases, shut down during kernel execution for aggressive power savings. Kernel software runs on the tile control core (CRV), while the Core Task Manager (CTM) bridges the CRV and the BE functional units and orchestrates tasks across a configurable set of units.

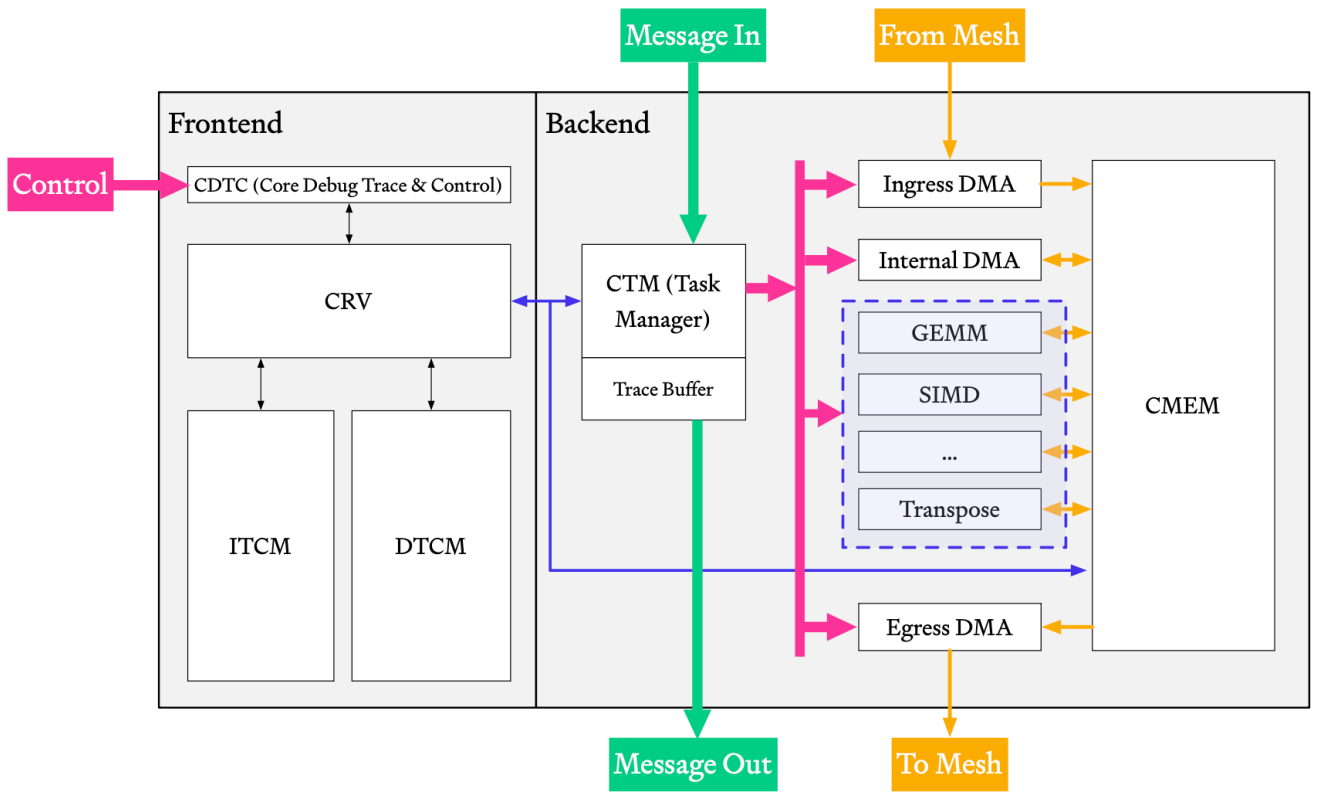

Fig. 2: Redwood tile architecture.

The hardware units within the tile BE are co-designed for modern transformer workloads, including on-device prefill and decode. The compute units include GEMV and GEMM engines for matrix operations built by arrayed Integer-based Multiply-Accumulate (MAC) units, and multi-lane SIMD engine for element-wise operations built by arrayed Floating-Point Units (FPU) capable of handling reductions, LUT-based operations, etc. (Fig. 3). One optimization uses the emulated-softmax algorithm from FlashAttention-4 [8], which reuses existing SIMD resources for an otherwise area-expensive operation. These functional units and the CRV share access to a local 512-KB core memory (CMEM) over a high-bandwidth bus. Local ingress and egress DMA engines move data into and out of CMEM.

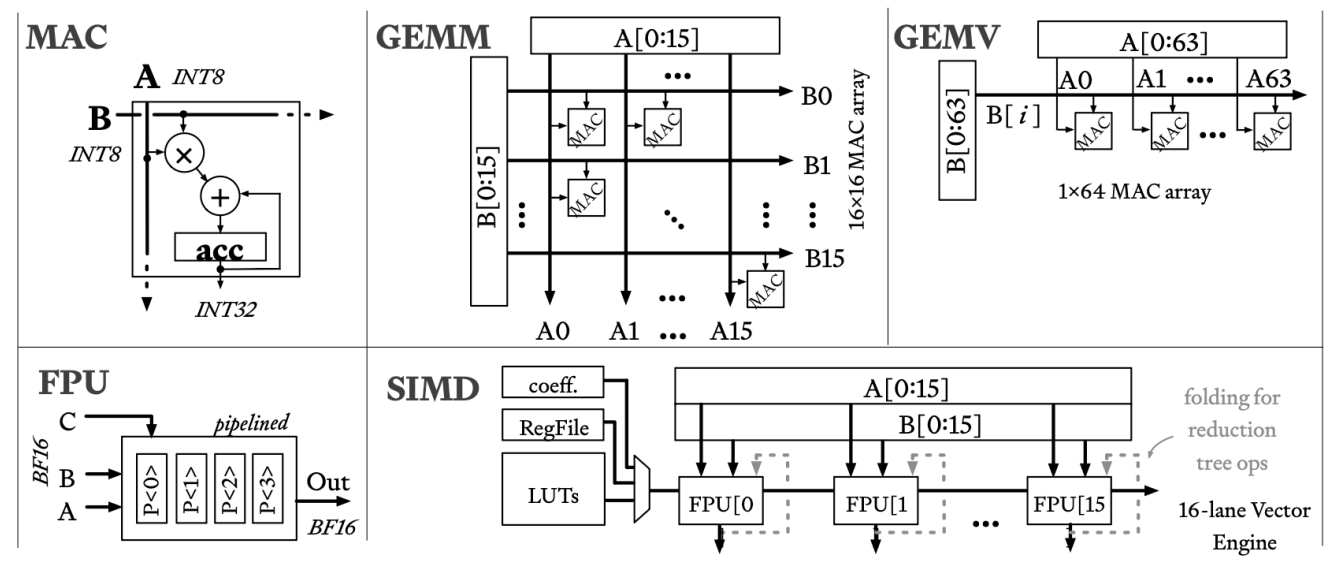

Fig. 3: Standard Redwood tile BE units.

### B. Control Mechanism

To compute using a Redwood tile, a kernel is first loaded into the FE instruction tightly coupled memory (ITCM). The Core Debug, Trace, and Control (CDTC) unit then boots the bare-metal tile control core (CRV). The CRV waits for a function call delivered to the data tightly coupled memory (DTCM) through the CDTC and then executes the selected kernel function. A kernel function typically expands into several MMIO writes that enqueue CTM tasks for dispatch to the corresponding functional units. Multiple kernel calls can be enqueued provided their implementations reside in ITCM. The overall FE-BE orchestration is illustrated in Fig. 4.

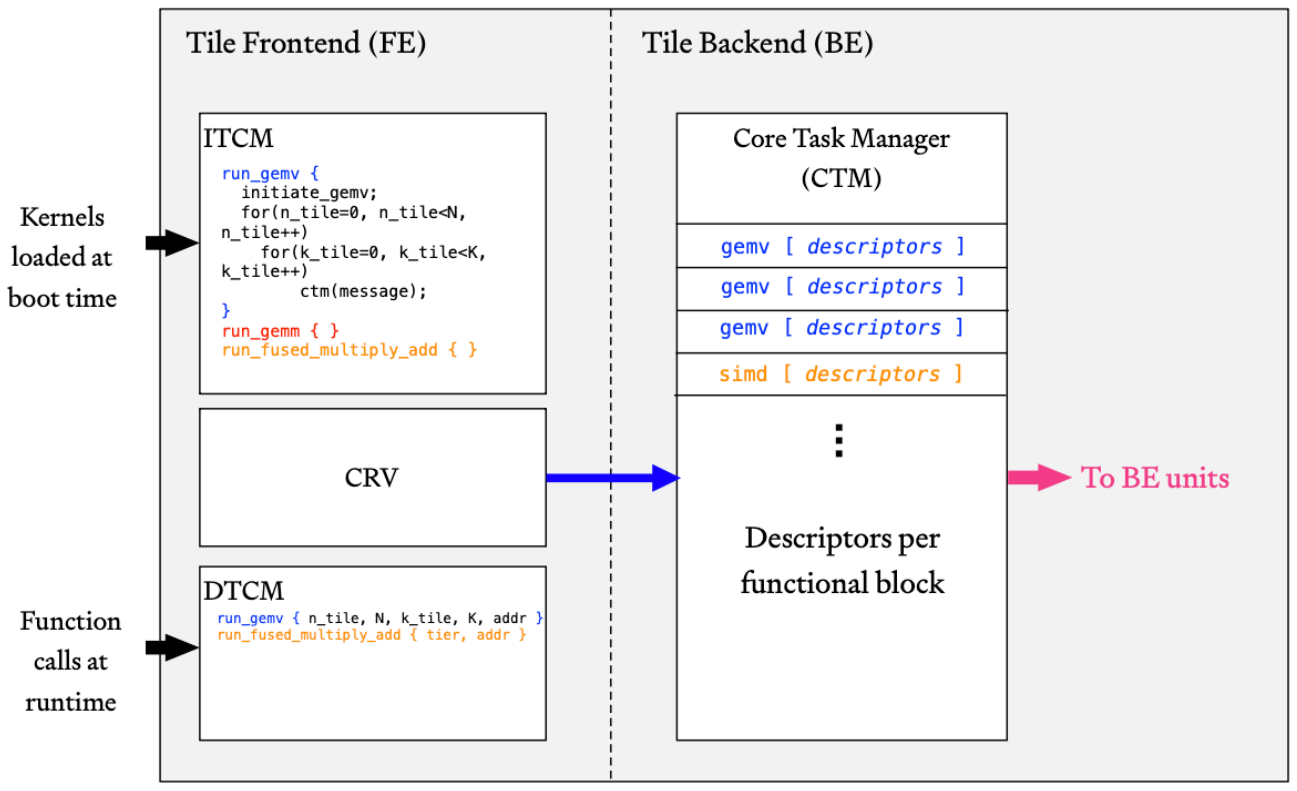

Fig. 4: Redwood FE-BE orchestration.

This decoupling between the FE and BE provides the following benefits:

- The CRV remains simple, implementing a minimal RISC-V specification with low area and power overhead.
- The CRV decoder remains unchanged when BE functional units are added, changed, or removed.
- The CRV can hand a task list to the CTM and then idle until interrupted.

The CTM natively supports:

- Arbitrary task ordering and fencing using task IDs to track out-of-order completions.

- Hardware tracing and logging to a trace buffer, with software notifications delivered through interrupts.
- Loops over arbitrary sections of queued tasks to reduce repeated CRV writes.

CTM tasks can also be fenced by external "messages" carried over the Redwood SoC message fabric, which connects all CTMs. The system-level messaging across cores and CTMs is shown in Fig. 5.

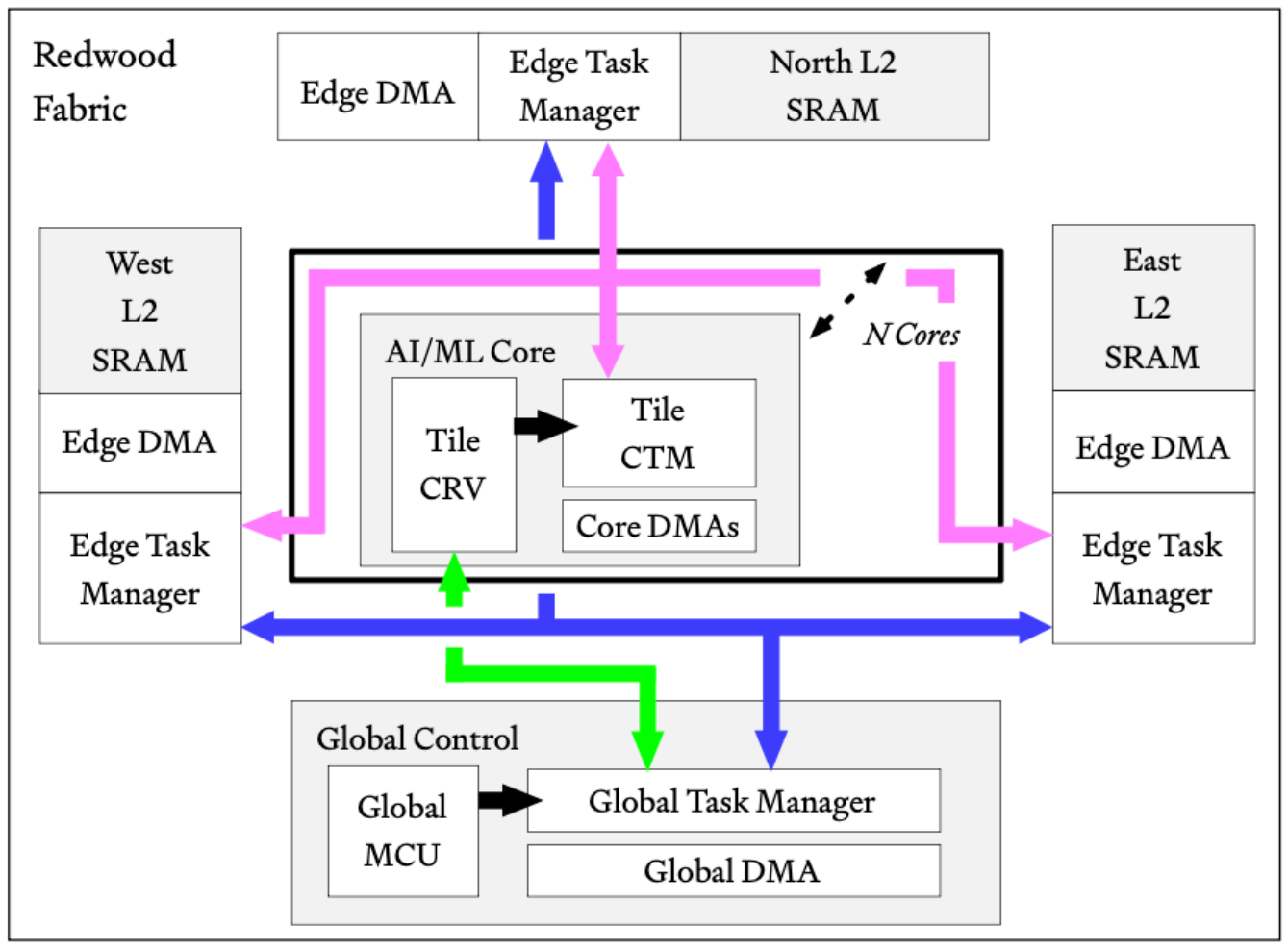

Fig. 5: Redwood CTM messaging network.

Messages allow CTMs to order control flow without involving the CRV or MCU. In the representative example in Fig. 6, the tile at (0,0) sends interleaved messages toward the West and East DMA engines, whose CTMs wait for "go-ahead" messages before releasing their next tasks. Communication can also flow in the opposite direction, with DMA CTMs signaling tile CTMs to send data downstream. Messaging can be configured as fire-and-forget or acknowledgment-based. The compiler uses messaging to coordinate prefetching, double-buffering, and out-of-order computation. Explicit traffic control through messages reduces the need for complex arbitration in the mesh and moves scheduling into the software stack.

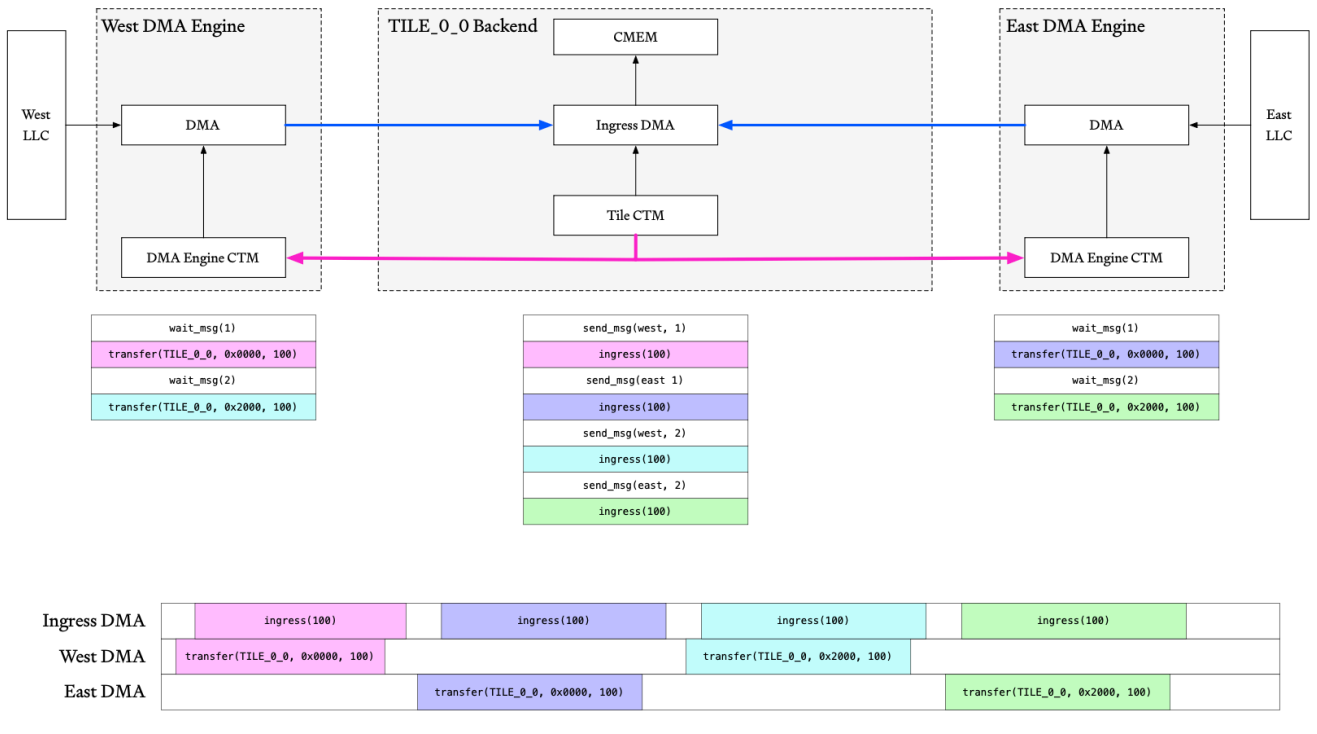

Fig. 6: Redwood CTM messaging flow example.

## III. Programming Model

Redwood includes a flexible programming model for running different models and workloads on the same fabric. Programs for the global control core (MCU) are called Dispatch Programs (DPs), while programs for individual tiles are called kernels. Multiple DPs are grouped into a "DP set," and multiple kernels into a "kernel set," amortizing initialization overhead.

For a host processor to use Redwood to perform an operation, the flow is as follows (Fig. 7):

1. (Prerequisite) Load the DP set from external memory into the MCU ITCM.
2. (Prerequisite) Load the kernel set from external memory into all tiles' ITCMs.
3. The host processor writes the dispatch ID and operands into the MCU DTCM and signals the MCU.
4. The MCU executes the dispatch program selected by the dispatch ID. The dispatch program can:
   - Program the static routing tables in the internal fabric mesh.
   - Configure the global task manager for prefetches, memory-to-memory copies, and scatter-gather operations.
   - Configure DMA-engine task managers to move data into and out of the tile array.
   - Launch tile kernels by writing a kernel ID and operands into a tile's DTCM and signaling the tile control core.

   The DP can poll status or rely on tile interrupts to determine when the tile array has finished. A running DP may launch one or more kernels over its lifetime. For example, FlashAttention repeatedly launches tiles to process different KV blocks and heads.
5. When the DP completes, the MCU interrupts the host processor to indicate that execution is complete and the reserved input and output buffers are available to the host.

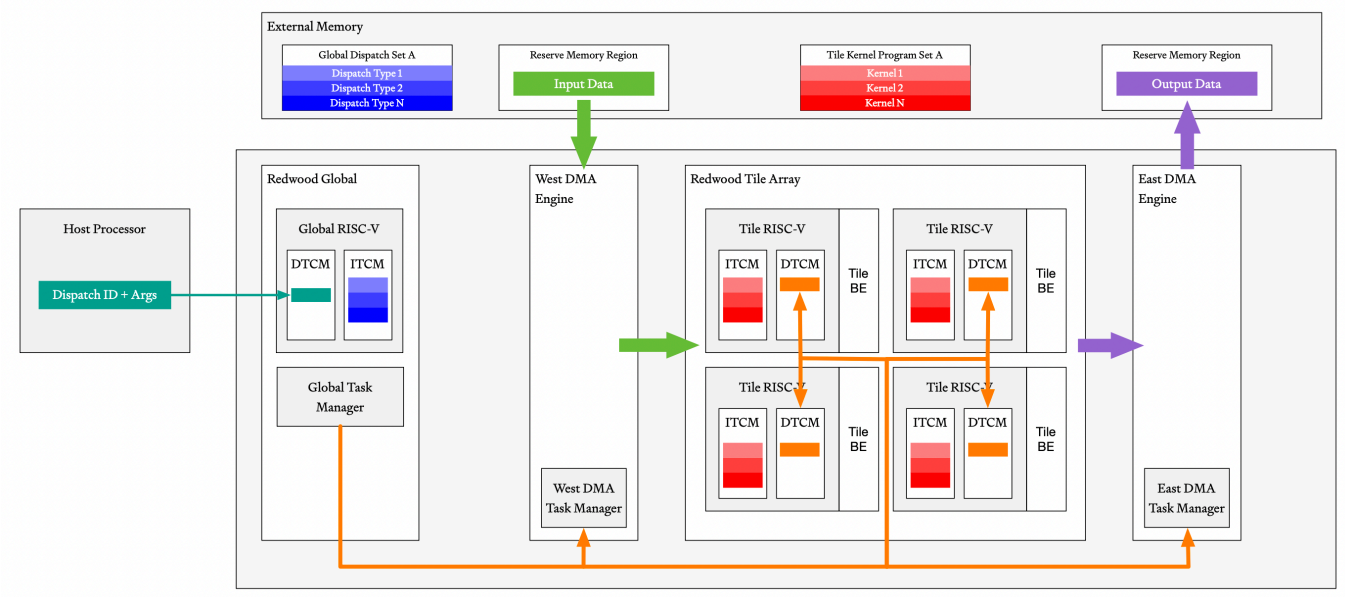

Fig. 7: Redwood dispatch and kernel execution flow.

If the DP or kernel sets required for an entire model do not fit in ITCM, the host processor loads them in partitions at runtime and groups them to minimize swaps.

## IV. Evaluation

Redwood Nano is the FPGA configuration of Redwood, designed and optimized for ultra-low-power, low-latency edge use cases. It consists of a $2 \times 2$ tile array, with each DMA engine connected to a set of 128-bit AXI4 interfaces. The West and North interfaces prioritize ingress read bandwidth, while the East interface prioritizes egress write bandwidth. Redwood Nano is synthesized and deployed on the AMD Versal VPK180 FPGA at 250 MHz (Fig. 8). To evaluate performance, we run Qwen3-0.6B on Redwood Nano and compare it with the NVIDIA Jetson Orin Nano. Fig. 8 shows the placement and instantiation hierarchy of Redwood Nano on the VPK180 FPGA.

We measure LLM decode performance in output tokens per second at both peak and average throughput. For Redwood Nano, the measurement includes sending the prompt from the host to the FPGA, running Qwen on the FPGA, and returning each output token to the host. The NVIDIA Jetson Orin Nano runs the same model at a 1020 MHz GPU clock, with performance measured using NVIDIA's Jetson WebUI. TABLE I presents the benchmark comparison.

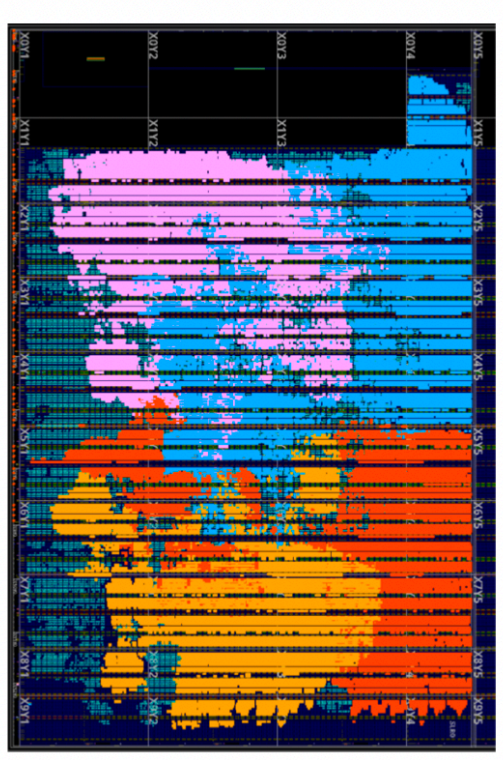

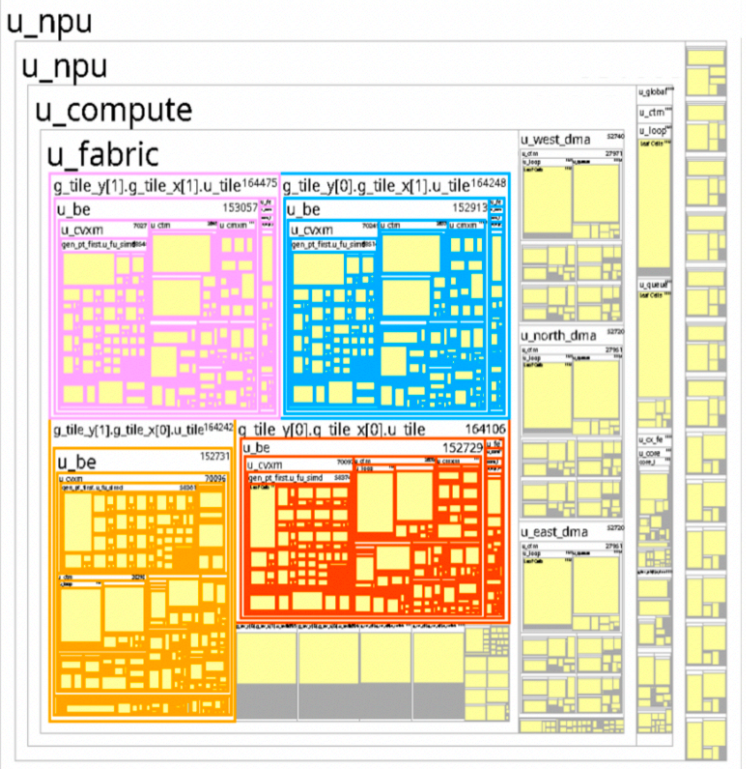

Fig. 8: Redwood Nano FPGA placement and instantiation hierarchy.

TABLE I: Redwood Nano vs. NVIDIA Jetson GPU performance comparison for Qwen3-0.6B LLM inference.

| Comparison | **Redwood** Nano | **NVIDIA GPU** Jetson Orin Nano |
|---|---|---|
| Average Tokens/s (*) | 12.1 | 28 |
| Frequency (MHz) | 250 | 1020 |
| Memory Bandwidth (GBytes/sec) | 16** | 68 |
| Memory Type | LPDDR4 | LPDDR5 |

*Measured over 128 tokens generated.
**DRAM Memory BW for Redwood at peak achievable frequency of 250MHz, 4x 128-bit AXI4 streams.

### A. Redwood Roofline Analysis of Peak Decoding Throughput

We first start our analysis by calculating the peak theoretical performance. The operation-level roofline is derived from the Qwen3-0.6B decode graph. The model has 28 decoder layers, hidden width 1024, intermediate width 3072, 16 query heads, 8 KV heads, head dimension 128, and vocabulary size 151,936. In each layer, the INT8 linear shapes are $Q : [2048, 1024]$, $K, V : [1024, 1024]$, $O : [1024, 2048]$, gate, up $: [3072, 1024]$, and down $: [1024, 3072]$.

The matrix engine contains four tiles with 64 INT8 MAC lanes per tile. Counting a multiply and accumulation as two operations, its peak rate is

$$P_{\text{GEMV}} = 4 \times 64 \times 250 \times 10^6 \times 2 = 128 \text{ Gop/s}.$$

The 512-bit SIMD datapath accepts 32 BF16 or 16 INT32 operands per cycle, giving peak rates of

$$P_{\text{SIMD,BF16}} = 32 \times 250 \times 10^6 = 8 \text{ Gop/s},$$

$$P_{\text{SIMD,INT32}} = 16 \times 250 \times 10^6 = 4 \text{ Gop/s}.$$

The sustained external-memory bandwidth is

$$B = 0.90 \frac{32 \times 3900 \times 10^6}{8} = 14.04 \text{ GB/s}.$$

For each unfused operator $i$, we count arithmetic work $O_i$ and total DRAM traffic $Q_i$, including all operand reads and result writes. Its compute, memory, and attainable times are

$$T_i^{\text{cmp}} = \frac{O_i}{P_i}, \quad T_i^{\text{mem}} = \frac{Q_i}{B}, \quad T_i = \max(T_i^{\text{cmp}}, T_i^{\text{mem}}).$$

Operators remain sequential because each consumes the preceding operator's result. Consequently, the complete-token bound is $T_{\text{token}} = \sum_i T_i$. Memory and compute may overlap within one operator, but no work or traffic is hidden across operator boundaries.

TABLE II: Operation-level roofline for one Qwen3-0.6B decoder layer at context length 128. Each row may group adjacent unfused operators; its attainable time sums the individual operator maxima.

| One-layer operator group | Work (Mop) | DRAM (MB) | $T_{\text{cmp}}$ (µs) | $T_{\text{mem}}$ (µs) | $\sum T_i$ (µs) | Limit |
|---|---|---|---|---|---|---|
| Input norm + QKV quant. | 0.008 | 0.009 | 1.0 | 0.7 | 1.0 | SIMD |
| Q/K/V projections | 8.397 | 4.247 | 68.6 | 302.5 | 303.2 | DRAM |
| Q/K norm, RoPE, and cache | 0.034 | 0.042 | 4.2 | 3.0 | 4.7 | SIMD |
| QK and score scaling | 0.530 | 0.158 | 6.1 | 11.2 | 12.3 | DRAM |
| Softmax and probability quant. | 0.029 | 0.039 | 3.6 | 2.8 | 4.3 | SIMD |
| V quantization and PV | 1.053 | 0.553 | 71.2 | 39.4 | 77.3 | SIMD |
| O projection and residual | 4.206 | 2.124 | 34.7 | 151.3 | 152.0 | DRAM |
| MLP norm and input quant. | 0.008 | 0.009 | 1.0 | 0.7 | 1.0 | SIMD |
| Gate and up projections | 12.595 | 6.367 | 102.9 | 453.5 | 454.6 | DRAM |
| SiLU and gate multiply | 0.022 | 0.031 | 2.7 | 2.2 | 3.6 | SIMD |
| Down projection and residual | 6.307 | 3.177 | 51.6 | 226.2 | 227.3 | DRAM |
| **One decoder layer** | **33.189** | **16.755** | **347.5** | **1193.4** | **1241.4** | **Mixed** |

One decoder layer therefore has an architectural latency of 1.241 ms (TABLE II). The 28 layers contribute 34.76 ms, while embedding, final normalization, the language-model head, and argmax contribute another 11.26 ms. The complete token moves 0.627 GB, requiring 44.65 ms of aggregate DRAM service and 12.29 ms of aggregate arithmetic service. Summing the per-operator roofline times gives

$$T_{\text{token}} = 46.02 \text{ ms}, \quad R_{\text{roof}} = \frac{1}{T_{\text{token}}} = 21.73 \text{ tokens/s}.$$

The memory service time is 3.6 times the arithmetic service time, establishing that decoding is strongly memory-bound, as expected. The measured throughput remains below this architectural ceiling because launch, synchronization, pipeline fill/drain, and host overhead are excluded. If memory transfer and execution are instead forced to serialize within every operator, the corresponding conservative bound is 56.94 ms per token, or 17.56 tokens/s. Further optimization remains in software and host overhead through better prefetching, software scheduling, and hardware updates in the control path.

### B. Redwood Performance, Area, and Power Projection

We estimate the architectural roofline for the configuration that uses a 1 GHz logic clock, which we believe to be reasonable given FPGA timing. The roofline model bounds throughput by the slowest stage along the weight-delivery path (DRAM → NoC → edge DMAs → mesh switches → compute) so we compute the sustainable bandwidth of each stage and take the minimum. A key point is that the DRAM data rate (3900 Mb/s per pin) is fixed by the memory device and does not scale with the fabric clock; the 1 GHz clock accelerates the on-chip fabric, SIMD engine, and MAC arrays. Assuming the same bandwidth as Nvidia Jetson Orin Nano, the tile-ingress path is still the narrowest stage, so the design remains memory-delivery-bound at approximately 64 GB/s. In this scenario, the architectural ceiling is approximately 95 tokens/s averaged over 128 generated tokens. The gain over today's approximately 21-token/s roofline comes from engaging three controllers and increasing the logic clock, which raises aggregate tile-ingress bandwidth from 16 to 64 GB/s and scales the compute engines by four. In a non-ideal setup, not all execution of the model can be overlapped with data transfer. We profiled the execution of Qwen on the FPGA and carefully scaled every step based on the improved clock frequency, higher memory bandwidth, and better task scheduling due to fewer hardware restrictions. Our most conservative projection shows that the current ASIC design will achieve approximately 49 tokens/s on average over 128 generated tokens without changes to the software stack.

We project the area and power of Redwood on a Samsung 8 nm-class process comparable to that used by NVIDIA Jetson Orin Nano. Physical area was estimated using a standard bottom-up gate-equivalent (GE) methodology tailored to Samsung 8 nm physical-design rules. Standard-cell counts were converted into GE units by weighting the 2 million combinational cells and 500,000 sequential registers by their average relative gate sizes. Their sum yields the raw standard-cell area, representing active silicon without interconnect gaps. We add 15% overhead for design-for-test logic. A 70% placement-utilization factor then provides space for metal routing, power-grid rails, and decoupling capacitors. Finally, we add 20% area overhead for clock-tree-synthesis repeaters, timing-closure cells, and perimeter structures. As such, Redwood Nano's projected Samsung 8 nm NPU block area is $\approx 2.88\text{mm}^2$.

We estimate total Redwood power by separating dynamic and static components. Core dynamic power follows $P_{\text{dyn}} = \alpha C V^2 f$, while core static power is represented by $P_{\text{stat}}$. At a target frequency of 1.0 GHz on Samsung 8 nm and a nominal core voltage of $V_{\text{core}} = 0.75V$, the 2 million logic cells and 500,000 flip-flops and 512KB SRAM per tile consume approximately 0.958 W of dynamic power under the average switching activity observed for Qwen3-0.6B. Static leakage is $P_{\text{stat}} \approx 0.07$ W. Including the remaining SoC and clock-management components brings the total chip-side power to $P_{\text{total}} \approx 1.335$ W. This number also aligns from FPGA-based testing, and also is an upper bound as we have not accounted for the savings from aggressive clock and power gating which the Redwood architecture natively supports.

Running the same application on Jetson, we achieved 28 tokens per second. Jetson's average power for its CPU and GPU cores is 2.59 W with fusion enabled. For fairness, both power estimates include host and accelerator compute but exclude memory controllers and other peripheral components. As a result, Redwood's performance and power projections on a comparable Samsung 8 nm-class process show a 1.75x performance improvement and a 1.9x power reduction, yielding a 3.4x performance-per-watt improvement while retaining significantly shorter design time and time to market. TABLE III summarizes the projected performance, area, and power of Redwood Nano compared with NVIDIA Jetson Orin Nano.

TABLE III: Redwood Nano vs. NVIDIA Jetson Orin Nano power, area, and performance for Qwen3-0.6B LLM inference on a Samsung 8 nm-class process.

| Comparison | **Redwood** Nano | **NVIDIA GPU** Jetson Orin Nano |
|---|---|---|
| Average Tokens/s * | 49 (1.75×) | 28 |
| Power (W) | 1.335 (1.9×) | 2.59 |
| Area (mm²) | 2.88 | NA |
| Tokens/s per Watt | 36.7 (3.4×) | 10.8 |

*Projected over 128 tokens generated (ASIC).

## V. Architect Labs System

In under two weeks, the entire Redwood was designed, verified, brought to synthesis and physical-design readiness, and deployed as the Redwood Nano FPGA configuration. The accelerator was produced from scratch, with zero pre-existing accelerator IP. This was enabled not only by training our own models, building agent harnesses, and building AI-native EDA tools, but also by a fundamental rethinking of the hardware-design flow from software to silicon. The traditional chip-design life cycle is highly sequential, progressing through stages from architectural definition to final tapeout. Along the way, collateral is "frozen," and changes are either made ad hoc or reserved for the next generation. Hardware teams pipeline development so that, after Version N is frozen, the corresponding team begins Version N+1 (Fig. 9). This enables large hardware companies to release new hardware at a 9–12 month cadence.

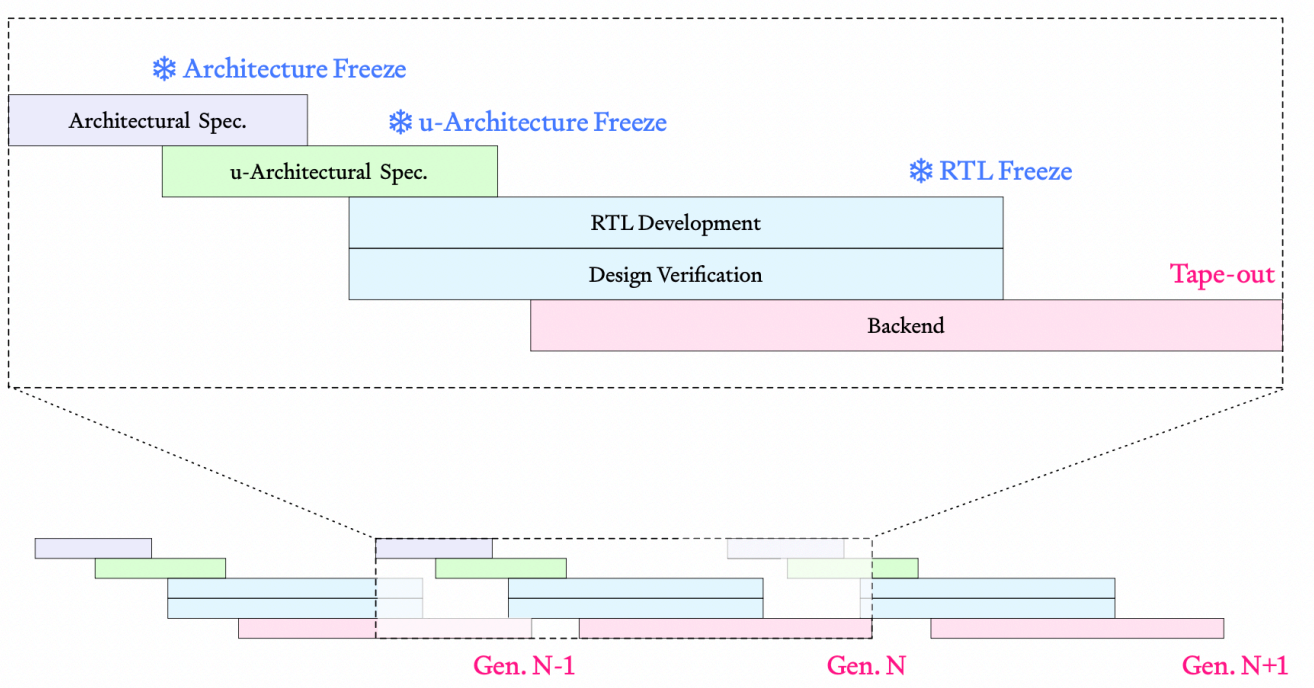

Fig. 9: Representative life cycle of a traditional ASIC program.

While its throughput may be acceptable, the latency of this sequential approach makes true hardware–software (HW–SW) co-design impractical. In the current AI landscape, by the time an architecture is defined and RTL design and verification are underway, new models can invalidate months of optimization. Hardware teams must therefore predict where workloads are heading and add general-purpose features as a hedge.

The Architect Labs flow is as automated and parallel as possible, removing the requirement for a "freeze" and enabling flexible architectural exploration and end-to-end implementation. It is built around the Architect Labs Platform (ALP), our in-house platform for end-to-end chip design (Fig. 10). Once design intent is captured in ALP, the automated flow generates RTL, UVM collateral, SVA assertions, formal proofs and other artifacts. No human intervention is required below the specification; human experts maintain ALP throughout the project life cycle and adjust the specification or design intent using functional, area, performance, timing, and power feedback.

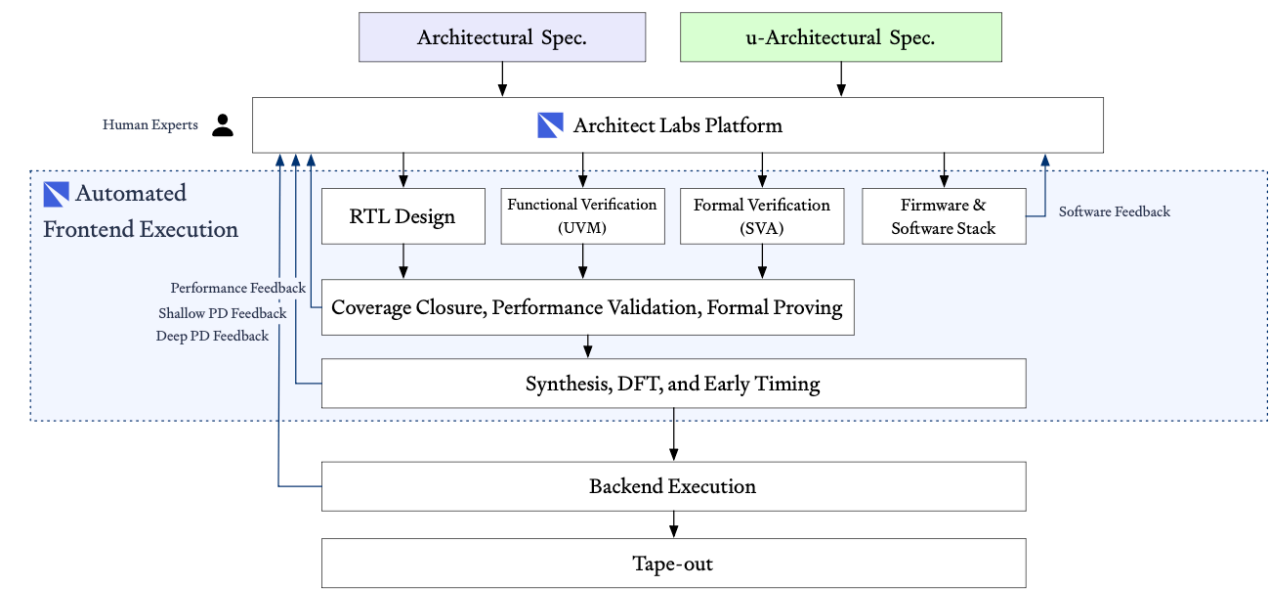

Fig. 10: Architect Labs Platform and end-to-end chip design flow.

A key benefit of this automated flow is that humans and agents can explore multiple architectural ideas in parallel. For Redwood, each architectural iteration was regenerated, reverified, and redeployed to FPGA within 48 hours. The complete design cycle, from the initial specification to finalized RTL, verification, firmware, custom kernels, and timing closure took two weeks, with all blocks reaching 95% code and functional coverage. A third week brought the target workloads i.e. new AI models onto the FPGA; the complete three-week project timeline is shown in Fig. 11.

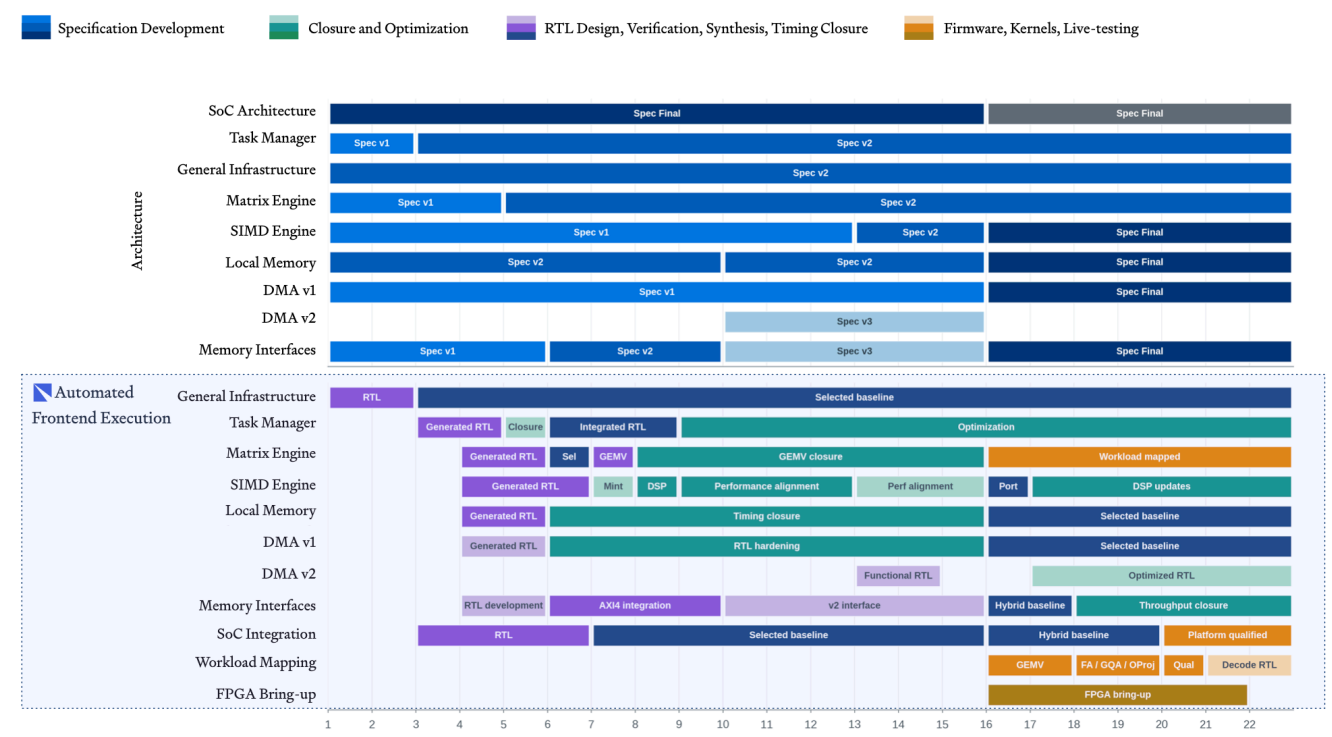

Fig. 11: Redwood design process history, extracted from repository commits.

Fig. 12 shows the repository's merge-commit history of the AI system over the project life cycle. The AI system reached a peak of 115 merge commits in one day while bringing target workloads online and continuously optimizing Redwood's firmware and kernels.

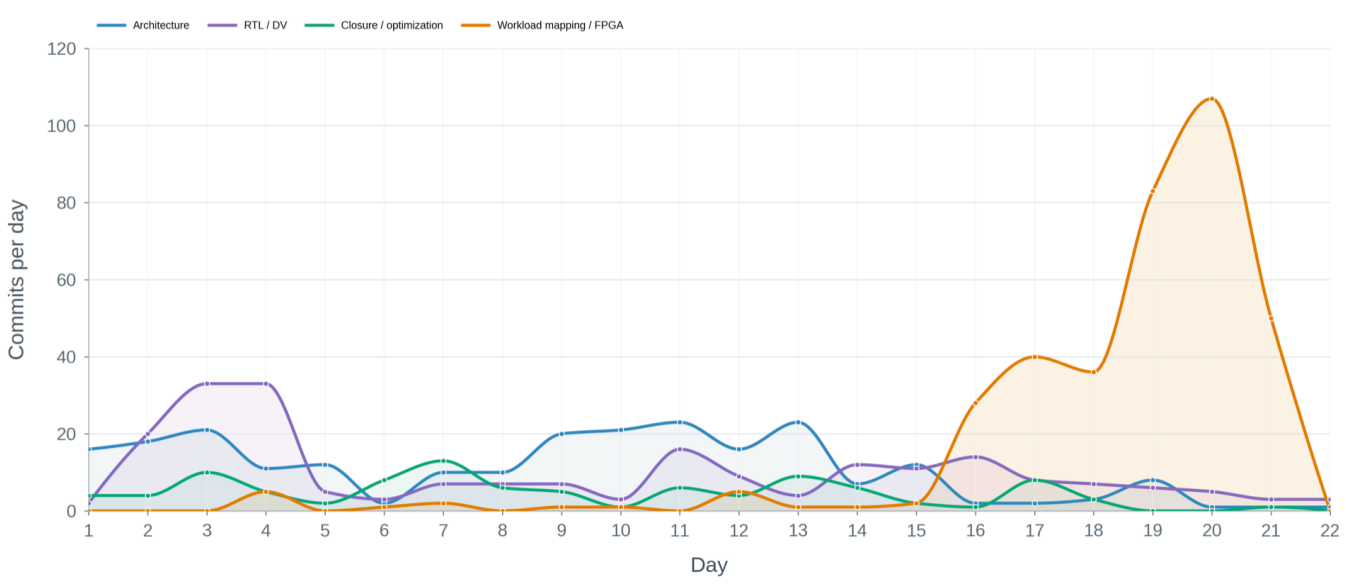

Fig. 12: Redwood repository commit activity.

## A. Automated Design Verification and Coverage Closure

Testbench generation, test-case development, and coverage closure for Redwood were fully automated. Traditional "AI agents for verification" often involve a human engineer using chat to automate test-case development and testbench construction. This approach still requires significant human effort and does not scale with chip complexity, so it has not materially shortened the end-to-end ASIC development cycle. In contrast, every Redwood testbench, test case, formal artifacts and simulation was generated automatically from ALP using AI and complier based methods without human DV participation.

Following standard industry verification practices, we use UVM technology alongside modern formal methods [9]. We have developed the first version of our proprietary formal engine to generate sections of each verification environment from a human-written specification. The verification flow measured and automatically optimized coverage and performance criteria. Every block achieved 95% code and functional coverage. When sending the first RTL drop from simulation to FPGA, there were *zero* bugs found, and to date our verification methods have not encountered an instance of a bug missed in our verification environment but revealed in actual hardware. As our technology scales, we expect verification rigor to scale with available compute rather than just the size of the team and EDA tool licenses, and we look forward to sharing more in future announcements.

## B. Design and Optimization

Because RTL design is fully automated in our flow, the system can explore a microarchitectural search space an order of magnitude larger than a human team can cover in the same time. Consider our SIMD engine, which performs reduced-precision floating-point and integer operations across multiple vector lanes. Although individual lanes are straightforward to develop and replicate, reduction operations such as reduce, max, and min span all lanes. By running more system instances over longer horizons, we can discover a broad set of novel candidates. In the example illustrated in Fig. 13 and Fig. 14, our AI system traversed the performance-area-timing search space over multiple days, designing, verifying, and optimizing while maintaining code coverage and verification rigor.

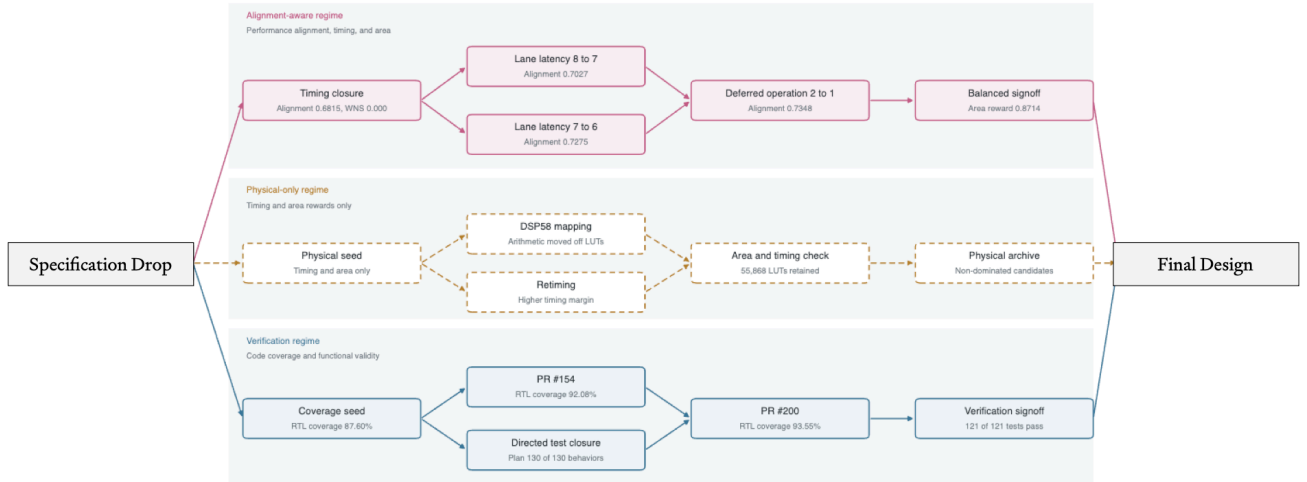

Fig. 13: Autonomous SIMD engine exploration path.

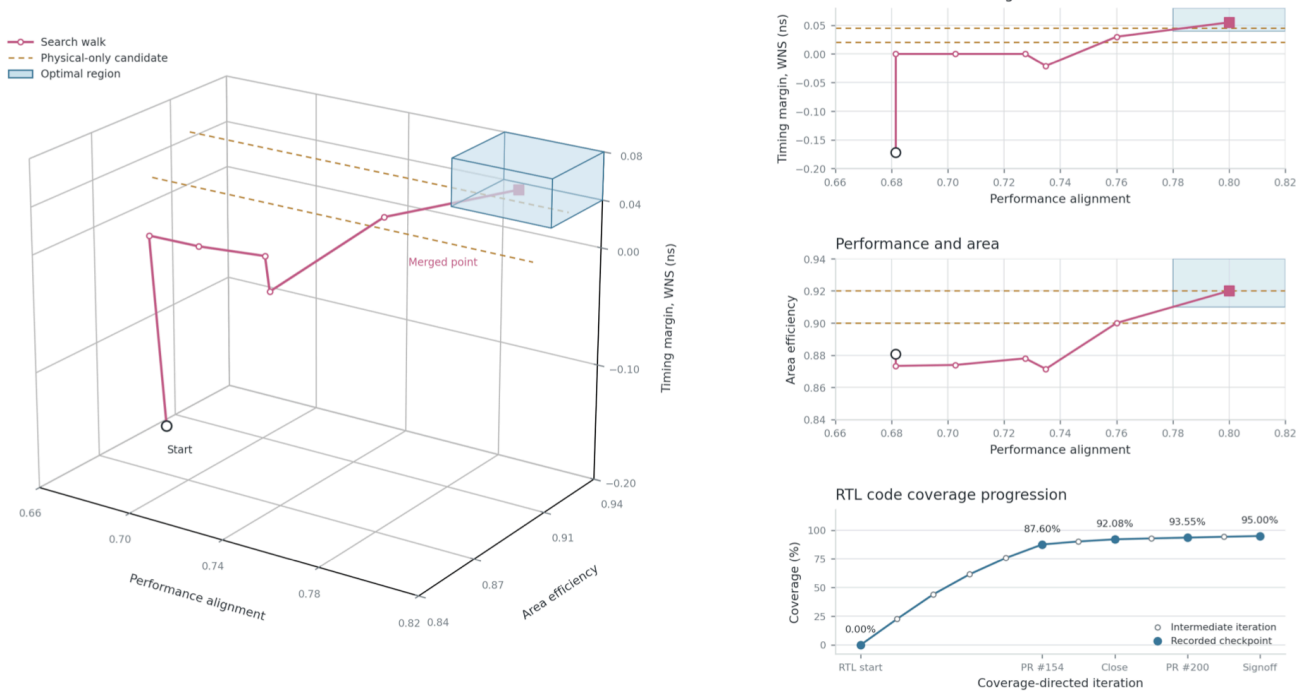

Fig. 14: SIMD engine micro-architectural exploration and optimization for performance, area, and timing, and code coverage.

Prior approaches to automated microarchitectural exploration were often limited to changes such as bit-width adjustments or register rearrangements. Here, the generated RTL candidates can use fundamentally different control paths, datapaths, and state machines and are free to find solutions beyond what humans may believe to be optimal. As our technology scales, we expect exploration quality to be bounded by available compute rather than by human insight, and as that bound rises, the system will surface architectures that lie outside the cognitive limits of the best human designers.

## C. Firmware Generation and Kernel Optimization

One benefit of ALP is that it enables co-development of all system software, including firmware, kernels, and performance models, before any RTL or verification collateral is written. This gives architects an unparalleled ability to make informed design decisions before committing to an implementation. Our AI system wrote and tested all firmware and kernels required to bring up the SoC and run Qwen inference. Depending on runtime, system software is tested against ALP projections, cycle-accurate RTL simulation, or an existing FPGA build. We designed an in-house custom emulation environment that multiplexes FPGA access across hundreds of concurrent agents, allowing them to run experiments, share performance results, and iterate without human intervention for days. In some instances, the AI system discovered optimizations and architectural improvements that our human experts had not considered or thoroughly explored. The system has also enabled us to discover new hardware features for the next-generation of Redwood, allowing multiple and parallel iterations of new designs.

Moving from RTL and performance-model simulation to our in-house FPGA environment reduced optimization runs from 15 hours to approximately 15–30 minutes. As AI automates more of the design process, we believe FPGA-based emulation will become essential for accelerating performance verification and ASIC roadmaps.

### D. Recursive Self-Improvement

As a final result, we deployed Qwen3 on Redwood and exposed it as an inference endpoint within our AI system. Through repeated sampling, the AI model discovered multiple timing improvements and kernel optimizations for several of its own operations, all at zero inference cost. We believe this is one of the earliest demonstrations of recursive self-improvement: an AI system designed an AI accelerator, deployed an AI model on it, and used that model to improve a future generation of the accelerator.

Today, there exists a gap between the capability of an AI system that can autonomously design hardware, and the capability of AI models that can actually be deployed on such hardware, as illustrated in Fig. 15. As our AI system continues to scale to more complex hardware designs, and we close this mismatch, sustained AI-driven improvements to modern hardware will become one of the largest drivers of AI progress. We believe this is the clearest measurement of recursive self-improvement.

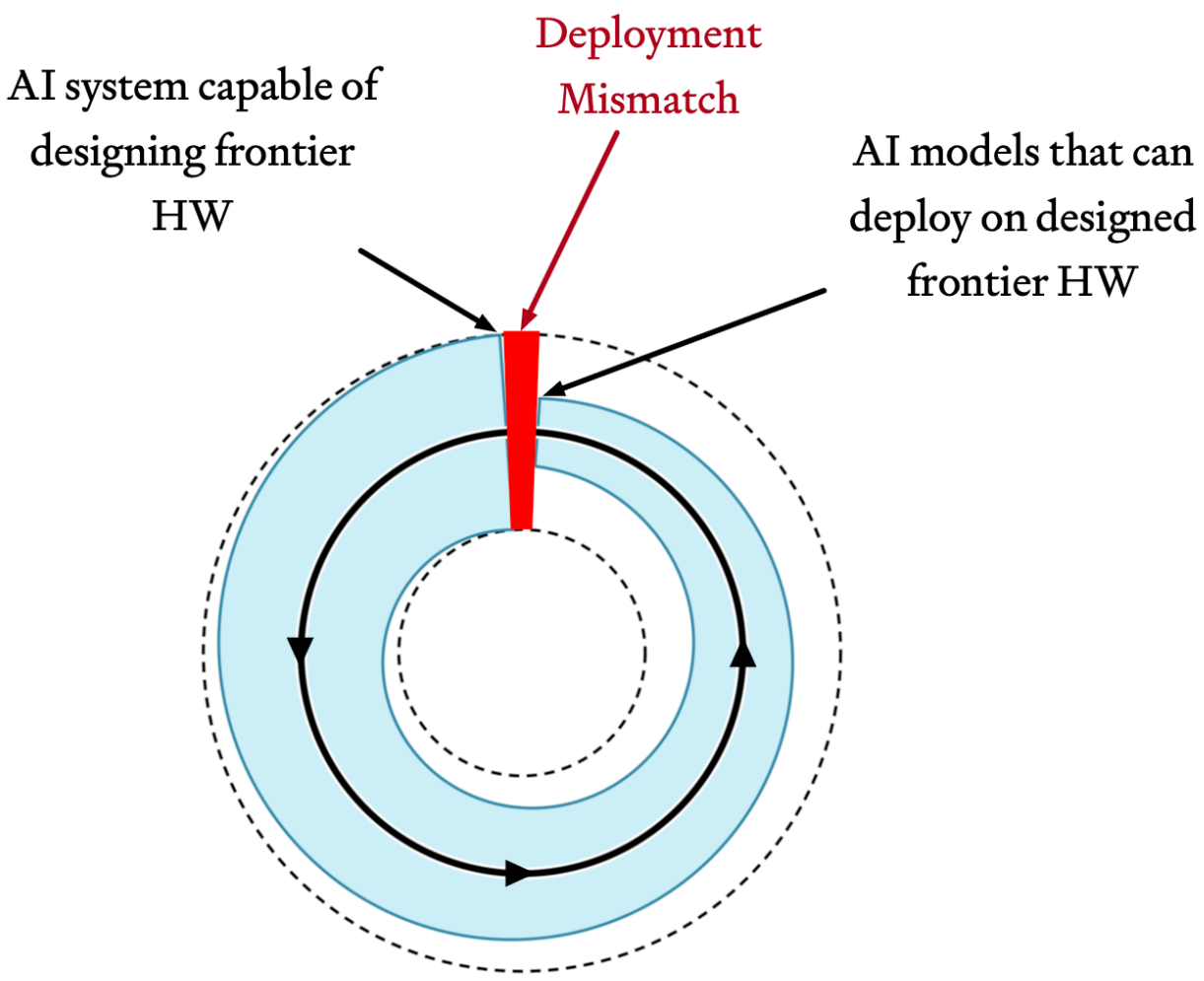

Fig. 15: Requirements for recursive self-improvement.

## VI. Conclusion

We present Redwood, a programmable accelerator designed, verified, and deployed from scratch by an AI system in under two weeks. The system closed every block at 95% code and functional coverage, and Redwood Nano ran Qwen3-0.6B on an AMD Versal FPGA at 13 peak and 12.1 average tokens/s. Calibrated by these FPGA measurements, our Samsung 8 nm-class projection reaches 49 tokens/s at 1.335 W, a $3.4\times$ performance-per-watt improvement over the measured Jetson Orin Nano baseline.

AI assistance within the conventional, sequential chip-design process has not substantially shortened end-to-end development. Our AI system takes an orthogonal approach: a high-level specification is the source of truth from which architecture, RTL, verification, firmware, and kernels are co-designed and optimized together. This allowed architectural changes to be reverified and redeployed to hardware in less than 48 hours.

Future work will focus on closing the remaining gap to the memory roofline, scaling Redwood to larger models and fabrics, and extending the AI system through physical design, tapeout, and post-silicon validation. Our ultimate goal is to make intelligence abundant through recursive self-improvement: as AI models and available compute improve, the system can explore and optimize more AI hardware designs, producing more efficient compute that further accelerates the next generation of AI and hardware development.

## Appendix: Contributors and Contact

Contributors, ordered alphabetically by last name: Armin Abdollahi, Vipin Boyanapalli, Trevor Daykin, Omar El Malki, James Fang, Joel Galenson, Dan Ganousis, Pin-Chun (Adrian) Hsu, Ebrahim Hussain, Moenes Iskarous, Arulmani Krishnan, Balbindra Kumar, Rakesh Kumar, Hugh Leather, Gang Li, Mu (Kevin) Lin, Madison Ma, Satyaveer Singh Mahecha, Basil Nabi, Deric Pang, Sahand Salamat, Aaditya Subedi, Ekin Sumbul, Raymond Wang, Manthan Wankar, Jindrich Zejda, Changyang Zeng.

General inquiries, including partnership & collaboration request, and technical questions about Redwood, may be directed to contact@architectlabs.com or architectlabs.com.